\documentstyle[aps,prl]{revtex}

\begin{document}

\twocolumn[\hsize\textwidth\columnwidth\hsize\csname %
twocolumnfalse\endcsname

\title{Preroughening, diffusion, and growth of a fcc(111) surface}
\author{Santi Prestipino$^{a,b}$, Giuseppe Santoro$^{a,b}$, and Erio
Tosatti$^{a,b,c}$}
\address{
a) Istituto Nazionale per la Fisica della Materia\\
b) International School for Advanced Studies, via Beirut 2-4, 34013 Trieste,
Italy\\
c) International Centre for Theoretical Physics, Trieste, Italy}

\date{\today}

\centering{
\medskip
\begin{minipage}{14cm}
{}~~~Preroughening of close-packed fcc(111) surfaces, found in rare gas solids,
is an interesting, but poorly characterized phase transition. We introduce
a restricted solid-on-solid model, named FCSOS, which describes it.
Using mostly Monte Carlo, we study both statics, including critical
behavior and scattering properties, and dynamics, including surface
diffusion and growth. In antiphase scattering, it is shown that preroughening
will generally show up at most as a dip. Surface growth is predicted to be
continuous at preroughening, where surface self-diffusion should
also drop. The physical mechanism leading to preroughening on rare
gas surfaces is analysed, and identified in the step-step elastic repulsion.
\end{minipage}
}

\maketitle


\vspace{5mm}

]
%%%%%%%%%%%%%%%%%%%%%%%%%%%%%%%%%%%%%%%%%%%%%%%%%%%%%%%%%%%%%%%%%%%%%%%%%%%%

Thermal disordering in the height profile of a crystal surface will
generally occur in two steps, as first shown by den Nijs. Using modified
solid-on-solid (SOS) models,\cite{denNijs89,denNijs92} he found that
regular roughening (temperature $T_R$) is preceded
at some $T_c<T_R$ by a critical, preroughening (PR) transition.
The intermediate disordered flat (DOF)
phase presents a very special form of surface disorder with proliferation
of up-down correlated steps which causes the first
layer to be only half-occupied. Based on the concept that PR is driven
by latent tendencies towards reconstruction,\cite{denNijs89,Mazzeo}
attempts at detecting PR focused mostly on metal surfaces,
with uncertain results.\cite{Conrad,Wang}
Surprisingly, the first convincing evidence of PR in unreconstructed
surfaces came instead from rare gases.\cite{YH,Day}
The (111) faces of Ar, Kr, and Xe, exhibit re-entrant layering in
adsorption isotherms above a $T_c\approx 0.85 T_m$ and below
$T_R\approx 0.95 T_m$ ($T_m$ being the melting temperature), with a
half-filled top layer between $T_c$ and $T_R$.
This is a clear indication of PR,\cite{Erice} and a simple hexagonal
model can in fact be constructed which contains PR with good similarities
with the experiment.\cite{Goodstein}

However, the driving force for PR on Ar(111) is certainly not reconstruction,
and remains to be understood. Moreover, the fcc(111) surface, with its three
sublattices, has a richer contents of degeneracy and symmetry than
simple hexagonal, which needs to be addressed. At this stage, in fact,
PR of fcc(111) surfaces is not well characterized at all.
We need, in particular, to understand scattering-related quantities
such as order parameters and susceptibilities, as well as important dynamic
properties like surface diffusion and growth.
This need is made more urgent by the strong possibility that PR (and not
roughening) might more generally constitute the true onset of surface
melting, as will be suggested by our surface diffusion results.

We have constructed a restricted SOS model of fcc(111) which
yields well-defined answers to these questions. We consider three
triangular sublattices $l=0,1,2$ with $ABCABC\ldots$ stacking. Heights in
the $l$-th sublattice are defined as integers $h_i=3n_i+l,\,\forall n_i$.
Nearest-neighbor height differences are constrained to be
$\pm 1,\pm 2$ (``infinite'' bond strength). A positive energy is
associated with height differences departing from the value attained
in the perfectly ordered fcc(111) surface (the ground state of the model).
The hamiltonian is:
\begin{eqnarray}
H &=& J\sum_{\langle 2\rangle} \delta (|h_i-h_j|-3)+K\sum_{\langle 3\rangle}
\delta (|h_i-h_j|-4) \nonumber \\
  &+& L\sum_{\langle 4\rangle} \delta (|h_i-h_j|-4)+M\sum_{\langle 4\rangle}
\delta (|h_i-h_j|-5)\,\cdots
\label{eq-1}
\end{eqnarray}
where $\sum_{\langle n\rangle}$ is shorthand for sum over all pairs of
$n$-th neighbors (distance measured in-plane), and
the couplings $J,K,L,M,\ldots$ are positive energy parameters.
We call this the FCSOS model (Fig.\ 1).
Anticipating the results, the minimal requirement in order to get a stable
DOF phase into the model is to have at least non-zero $K$ {\em and} $L$.
Longer range interactions, including $M$ and beyond, do not bring further
changes. In this sense the FCSOS with $L\ne 0$ is generic.

We work out the essential features of the FCSOS phase diagram by the strip
transfer-matrix method. Height parities $(-1)^{h_i}$ are chosen
to represent the SOS configurations of a $N\times\infty$ strip.\cite{BH}
The fairly long-range interaction $L$ obliges to compute all possible
states of {\em three} spin rows.
Our maximum strip size is $N=12$.
We locate phase boundaries through the vanishing of the free energy of
relevant interfaces.\cite{denNijs89}
Strictly {\em at} PR, we find that {\em all} off-plane excitations are
costless.
The step free energy vanishes at PR and remains zero inside the DOF
phase (due to proliferation of steps), whereas the cost of two parallel
steps is non-zero until roughening.
Within the transfer matrix framework these quantities are
evaluated by imposing suitable boundary conditions to the strip.
Details of the calculation are reported elsewhere.\cite{Santi}

We list a number of results (all the couplings from $M$ onward
are put equal to zero). When the only non-zero coupling in
the model is $J$, there is no PR, and a simple roughening is found at
$e^{\beta J}\simeq 1.6$. The situation remains unchanged when $K>0$
but $L=0$. In particular, roughening takes place at $e^{\beta J}\simeq 1.5$,
when $K=+\infty$. Finally, a PR transition is found along the
$K=L=+\infty$ line at $e^{\beta J}\simeq 1.5$. When $L=K$
and $J=0$ the boundary between the DOF and the rough phase is found at
$e^{\beta K}\approx 2$. The overall phase diagram is sketched in Fig.\ 1.
It resembles that of Ref.\ \cite{denNijs89}, and is in fact more general
(for positive couplings) as far as the nature of the phases it includes.

Monte Carlo (MC) simulations confirm the above findings. We use
$N\times N$ cells of increasing size, ranging from $N=24$ to
$N=96$, and a standard Metropolis algorithm. After equilibration,
more than two million MC sweeps are produced and average quantities
such as the interfacial width $\langle\delta h^2\rangle$, specific heat,
and parity order parameter
$P=(3/N^2)\langle\sum_{i}(-1)^{h_i}\rangle\,,$ are evaluated.
In an ideal DOF configuration the average height is half-integer,
$\delta h^2=11/12$ due to half-occupancy in the topmost layer,
$P$ should be zero, and its susceptibility $\chi_P$ should diverge at PR.

As Fig.\ 2 shows, on the $K=L=+\infty$ line, $P$ indeed vanishes near
$e^{\beta J}=1.5$ and $\chi_P$ diverges. In the
DOF phase, the average height is half-integer (Fig.\ 3), and correspondingly
the occupancy of the top layer is one-half. Moreover,
both the specific heat and its derivative (not shown) remain finite at
$T_c$, thus signalling a higher-order transition than Ising, or 3-state
Potts. The ordered phase and the DOF phase have {\em the
same degeneracy} (here 3-fold). As a consequence, PR is expected to be
non-universal\cite{denNijs89} unlike Ising or 3-state Potts,
where the degeneracy is fully removed above $T_c$.

Since the free energy of all step excitations is zero at PR,
one expects the height fluctuations to diverge at $T_c$,
$\langle\delta h^2\rangle\sim\frac{1}{4\pi K_c}$ln$N$,
but {\em not below or above} (until $T_R$).
The MC results in Fig.\ 2 support this.
Because of the Gaussian behavior at $T_c$,
critical exponents along the PR line can be recovered in terms of
one parameter only, namely $K_c$.\cite{denNijs89,Erice}
At $K=L=+\infty$ and $e^{\beta J}=1.5$ we estimate $K_c\approx 1.06$,
whence $\eta=\frac{\pi}{4K_c}\approx 0.74$,
$\nu=\frac{2K_c}{4K_c-\pi}\approx 1.92$,
$\alpha=2-2\nu\approx -1.85$,
$\beta=\frac{\pi/4}{4K_c-\pi}\approx 0.71$, and
$\gamma=2-\alpha-2\beta\approx 2.42$.
At finite $L$ and $K$, the PR line is nearly insensitive to $L/K$,
while roughening (where $\langle\delta h^2\rangle\sim\pi^{-2}$ln$N$)
shifts to higher $\beta K$ as $L/K$ drops. Hence, the PR temperature
is only controlled by the $J$ value, which is then about $0.25$ for argon
($T_c\approx$ 70 K),\cite{YH} in units of the Lennard-Jones $\epsilon$
(120 K for Ar).

PR can be revealed in scattering
experiments.\cite{denNijs89,denNijs92,Mazzeo,Bernasconi}
Antiphase elastic X-ray or atom scattering is given by
\begin{equation}
I\left({\bf Q},q_z=\frac{\pi}{a_z}\right)\propto
S^2\delta_{{\bf Q},{\bf G}}+\frac{k_BT}{N^2}\chi_S({\bf Q})\,,
\label{eq-3}
\end{equation}
where $a_z$ is the vertical layer separation, ${\bf Q}$ and $q_z$ are the
surface parallel and perpendicular momentum transfer, respectively.
Here $S=(3/N^2)\langle\sum_{i}\alpha_i(-1)^{h_i}\rangle$ contains
a shadowing factor $\alpha_i$,\cite{Mazzeo} which may be taken, for
example, to be one
for local surface maxima and zero otherwise, and $\chi_S({\bf Q})$ is
the local susceptibility. We find (Fig.\ 2) that the Bragg amplitude $S^2$
has a dip and vanishes at PR. However, unlike $P$, it
is non-zero on both sides of the transition point (Fig.\ 2).
This experimental signature can be used to detect PR on surfaces,
including metals.

%
%         DYNAMICS
%
Surface dynamics near PR is so far unexplored. We consider
first surface growth. Continuous growth in the DOF phase is unlikely since
the free energy of two parallel steps is non-zero. This implies existence
of a free energy barrier for the formation of a stable growth nucleus,
and a behavior $m\propto e^{-C/\Delta\mu}$ for the growth mobility,
where $\Delta\mu$ is the overpressure (smaller than a threshold $\Delta\mu_c$)
and $C$ is proportional to the square of the cost of two parallel steps.
At finite size, we can also learn indirectly about the surface mobility
$m(N)$ from the behavior of the average surface height $\bar{h}(t)$
at equilibrium.
The surface {\em as a whole} diffuses in the form $\langle(\bar{h}(t)-
\bar{h}(0))^2\rangle\sim 2d(N)t$, with
Einstein-like proportionality between $d(N)$ and $N^{-2}m(N)$.
Fig.\ 3 shows both the equilibrium evolution of $\bar{h}$ (left), and the
growth behavior at finite $\Delta\mu$ (right). At equilibrium
the surface behaves like a Stokes particle
undergoing Brownian motion. The height is quantized, and is integer
below $T_c$ and half-integer above $T_c$.
At PR quantization disappears, and $d(N)$ is clearly much larger.
This behavior is fully confirmed when $\Delta\mu >0$. In both flat
and DOF phases, growth is activated and occurs between quantized
levels. At $T_c$ the surface is rough, and $m$ soars accordingly,
indicating continuous growth.

The next dynamical issue is single-particle surface diffusion near the
PR transition.
On account of the critical slowing-down affecting dynamical processes at a
continuous transition, and of a finite coupling of a migreting partucle to
the DOF order parameter, we should expect a drop of the diffusion
coefficient at PR.\cite{Ying}
Dynamic scaling hypotesis combined with the assumption of Gaussian spreading
of density inhomogeneities out of criticality predicts that diffusion should
vanish as $D\sim\mid t\mid ^{\frac{\gamma (z-2)}{2-\eta}}\,,$
where $z$ is a dynamical critical exponent and $t=\frac{T-T_c}{T_c}\ll 1$.
In the case of PR, the rugged landscape is expected to concur
substantially to hinder particle hopping.
The tracer diffusion $D=\langle\Delta r^2\rangle /4t$ is extracted
from a particle-conserving (Kawasaki) MC simulation and
displayed in Fig.\ 4. The size-dependence is maximum near $T_c$,
indicating a drop in the thermodynamic limit, as expected.
A similar drop is
found when approaching roughening from the DOF side (not shown).

We are now ready to make contact with real fcc(111) surfaces, such
as Ar(111). Modelling Ar by a Lennard-Jones 12-6 potential
($\sigma=r_{\rm NN}/1.0933$, $\epsilon=120K$, cut-off at $r=3.2\sigma$),
we consider a variety of step geometries, and use
simulated annealing to optimize the $T=0$ geometry and energy.
For both (100) and (111) single steps we find an energy
(units of $\epsilon$) $\simeq 0.58$ per unit lenght.
This $J$ value is about twice as large as that required
for a PR transition at $70K$. This is not
surprising at all, in view of the additional vibrational and free-volume
reduction factors which must be applied to $T=0$ step energies
before using them near $T_m$. We further find that two adjacent
steps cost 1.14 (at the shortest distance) if
up-down, and 1.49, 1.35, 1.22, $\ldots$,
$2\cdot 0.58+\alpha/(l-l_0)^2$ ($\alpha/r_{\rm NN}^2\approx 1.0,
l_0\approx -0.5$), if up-up at
distances $1/2, 1, 2, \ldots ,l$ lattice spacings, respectively.
The lack of up-down interaction in FCSOS is thus justified.
The relatively short-range up-up repulsion needs to be qualified.
To do this we determine effective FCSOS $K$ and $L$ values (we take
for simplicity $K=L$) so as to generate a probability distribution for
the relative distance of two parallel steps
which is closest to that generated by the true long-range repulsion,
using a method similar to Bartelt {\em et al}.\cite{Bartelt}
For instance, when two parallel steps are, on average,
$4, 4.5, 5$ lattice spacings apart, the long-range interaction
$\alpha /(l-l_0)^2$ enhances $K$ and $L$ from bare values of
$\approx 0.62$ and $\approx 0.16$, up to
larger effective values $K=L=2.0, 14.5,\,>10^4$, respectively.
With reference to Fig.\ 1, we conclude that long-range elastic up-up
step repulsion is crucial in giving rise to PR.
This has a very transparent physical interpretation.
Without step-step interactions ($K=L=0$), PR and roughening initially
coincide. Parallel-step repulsion leaves PR unaffected, but not roughening
which is pushed at higher temperature.

Now, some experimental consequences.
The re-entrant behavior seen in rare-gas absorption isotherms, with
half-filled layers above PR, agrees very well with our growth results
(Fig.\ 3). Continuous growth strictly at $T_c$ is found,
a remarkable feature of this system. We moreover find that antiphase
scattering, if feasible, should show the critical drop at $T_c$
predicted for $S^2$, followed by a recovery in the DOF phase,
as in Fig.\ 2. This finite value of $S^2$ is generic, so long as
the shadowing factor $\alpha_i\neq 1$.
Tracer diffusion along the surface, if measurable
on a time scale short enough with respect to evaporation events,
should also show a dip at $T_c$, providing evidence for entanglement
with critical fluctuations at PR.
Finally, since parallel-step elastic repulsion is
universal, we can expect that PR should also be common.
In fact, we surmise that the abrupt
vacancy proliferation generally found at the early onset of surface
melting \cite{Frenken} might be precisely related to PR.
It is hoped that these results will stimulate newer experimental efforts.

We acknowledge discussions with G. Jug and A. C. Levi.
This research was supported by the Italian C.N.R. under the
``Progetto Finalizzato `Sistemi informatici e Calcolo parallelo'\,''
and under contract 94.00708.CT02 (SUPALTEMP).

%
% FIGURE CAPTIONS
%
\newpage
\begin{center}
{\bf Figure Captions}
\end{center}

\begin{description}
\item[Fig.\ 1 ]
Phase diagram of the FCSOS model (inset) for $L=K$.
Thermal behavior of Ar(111) corresponds qualitatively to the line $K\gg J$.

\item[Fig.\ 2 ]
Preroughening critical behavior of the FCSOS model obtained by Monte Carlo
along
the line $K=L=+\infty$. From top to bottom:
parity order parameter $P$, susceptibility $\chi_P$,
Bragg scattering amplitude $S$, susceptibility $\chi_S$,
and interface width
$\langle\delta h^2\rangle =(1/N^2)\langle \sum_i (h_i-\bar{h})^2\rangle$.
Data for $N=18$($\bullet$), $24$($\triangle$), $36$($\Box$),
$48$($\circ$), $60$($\ast$), $72$($\times$), $96$(solid $\triangle$)
are shown. Dotted lines are guides to the eye.

\item[Fig.\ 3 ]
Average surface height $\bar{h}$ during a MC run, for $N=36$, at three
different temperatures along the $K=L=+\infty$ line.
Left: equilibrium ($\Delta\mu =0$). Right: growth ($\Delta\mu =0.0333$).
{}From top to bottom, $e^{\beta J}=1.3$ (DOF phase), $1.5$ (PR), and $1.7$
(ordered flat phase). Note the continuous growth at PR, and the half-integer
quantization in the DOF phase. Labels A, B, C describe the top layer
(half-layer in DOF) sublattice.

\item[Fig.\ 4 ]
Single-particle surface diffusion in FCSOS ($K=L=+\infty$).
Data for $N=18$($\bullet$), $36$($\Box$),
$48$($\circ$), $60$($\ast$), $96$(solid $\triangle$)
are shown. Note the strong decrease with increasing size,
suggesting a dip at PR. The dotted line is a guide to the eye.

\end{description}


\begin{thebibliography}{99}

\bibitem{denNijs89}
M. den Nijs and K. Rommelse, Phys.\ Rev.\ B {\bf 40}, 4709 (1989).

\bibitem{denNijs92}
M. den Nijs, Phys.\ Rev.\ B {\bf 46}, 10386 (1992).

\bibitem{Mazzeo}
G. Mazzeo {\em et al.\/}, Surf.\ Sci.\ {\bf 273}, 237 (1992);
Europhys.\ Lett.\ {\bf 22}, 39 (1993);
Phys.\ Rev.\ B {\bf 49}, 7625 (1994).

\bibitem{Conrad}
Y. Cao and E. Conrad, Phys.\ Rev.\ Lett.\ {\bf 64}, 447 (1990).

\bibitem{Wang}
H. N. Yang {\em et al.\/}, Europhys.\ Lett.\  {\bf 19}, 215 (1992).

\bibitem{YH}
H. S. Youn and G. B. Hess, Phys.\ Rev.\ Lett.\ {\bf 64}, 918 (1990).

\bibitem{Day}
P. Day {\em et al.\/}, Phys.\ Rev.\ B {\bf 47}, 7501 (1993);
Phys.\ Rev.\ B {\bf 47}, 10716 (1993);
H. S. Youn, X. F. Meng, and G. B. Hess, Phys.\ Rev.\ B
{\bf 48}, 14556 (1993).

\bibitem{Erice}
M. den Nijs, in {\em Phase Transitions in Surface Films} 2,
Eds.\ H. Taub et al. (Plenum, New York, 1991).

\bibitem{Goodstein}
P. B. Weichman, P. Day, and D. Goodstein, Phys.\ Rev.\ Lett.\
{\bf 74}, 418 (1995).

\bibitem{BH}
H. W. J. Bl\"{o}te and H. J. Hilhorst, J.\ Phys.\ A {\bf 15}, L631
(1982).

\bibitem{Santi}
S. Prestipino and E. Tosatti (unpublished).

\bibitem{Bernasconi}
M. Bernasconi and E. Tosatti, Surf.\ Sci.\ Rep.\ {\bf 17}, 363 (1993).

\bibitem{Ying}
T. Ala-Nissila, W. K. Han, and S. C. Ying,
J. Electron.\ Spectrosc.\ Relat.\ Phenom.\ {\bf 54-55}, 245 (1990);
P. C. Hohenberg and B. I. Halperin, Rev.\ of Mod.\ Phys.\ {\bf 49}, 435
(1977).

\bibitem{Bartelt}
N. C. Bartelt, T. L. Einstein, and E. D. Williams, Surf.\ Sci.\ Lett.\
{\bf 240}, L591 (1990).

\bibitem{Frenken}
A. W. Denier van der Gon {\em et al.\/}, Surf.\ Sci.\ {\bf 256}, 385
(1991).

\end{thebibliography}
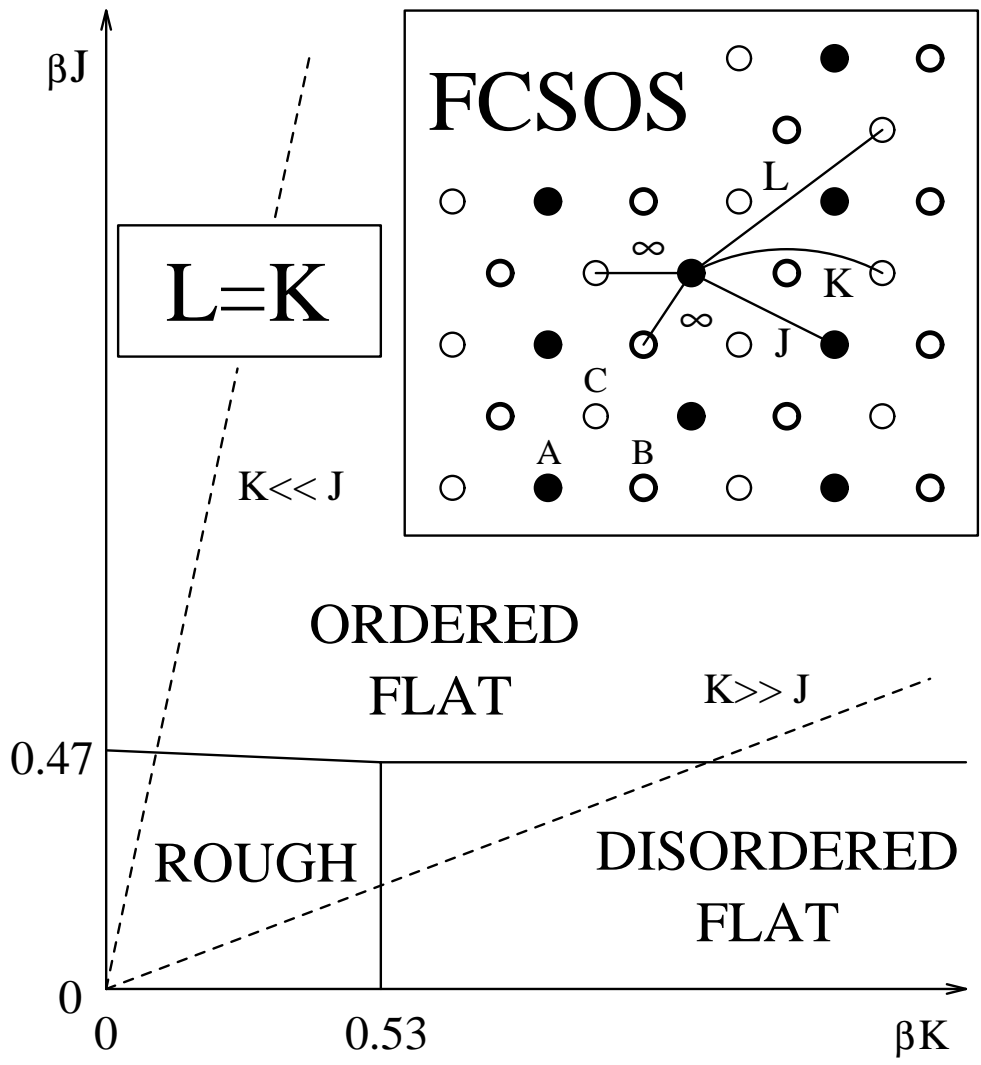
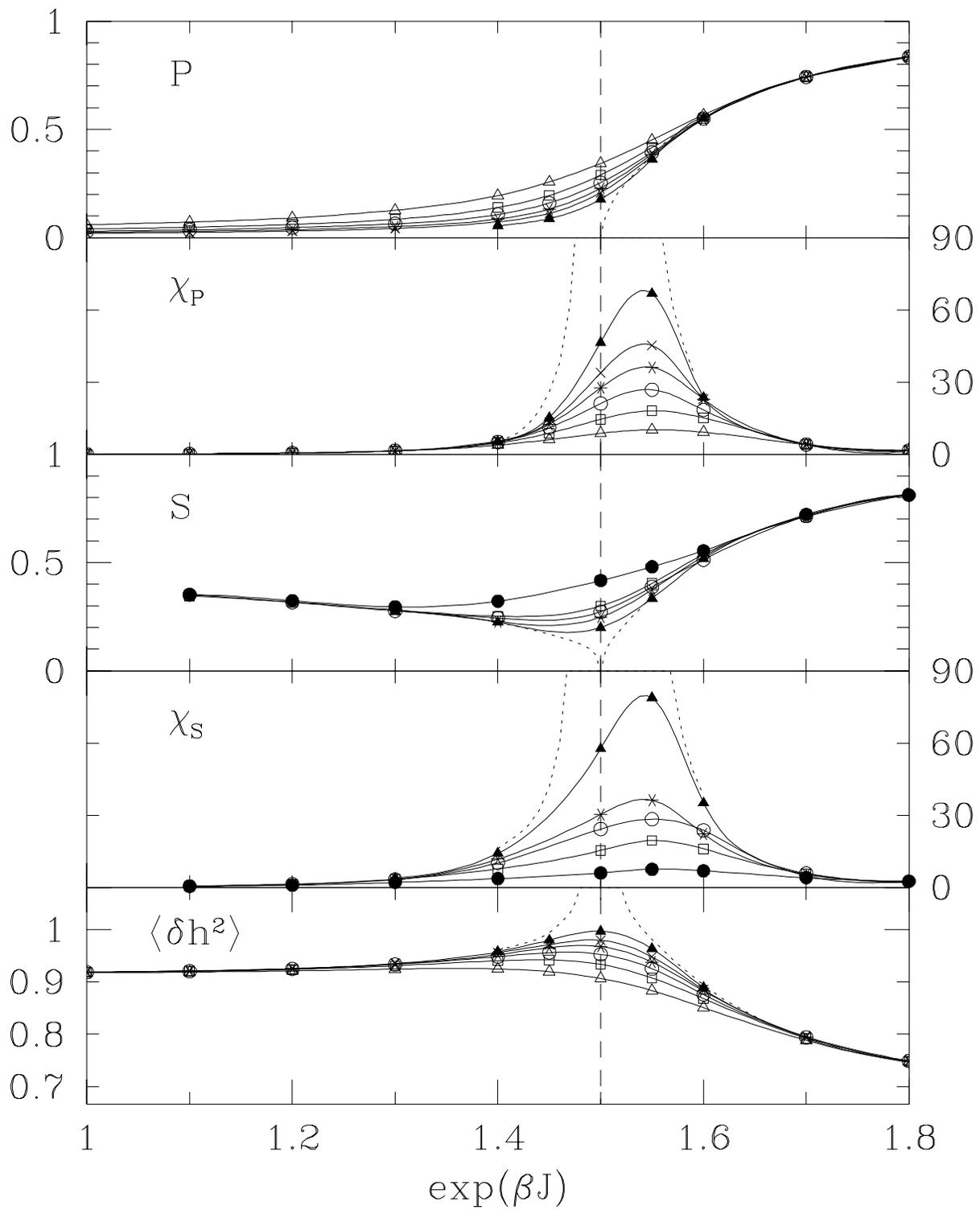
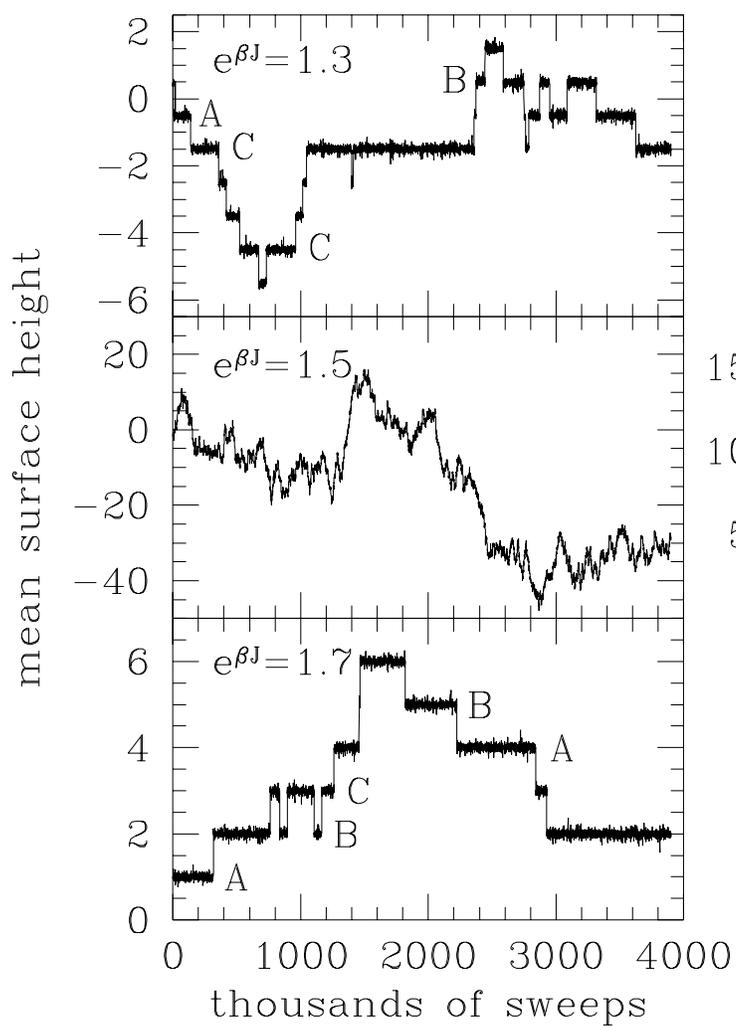
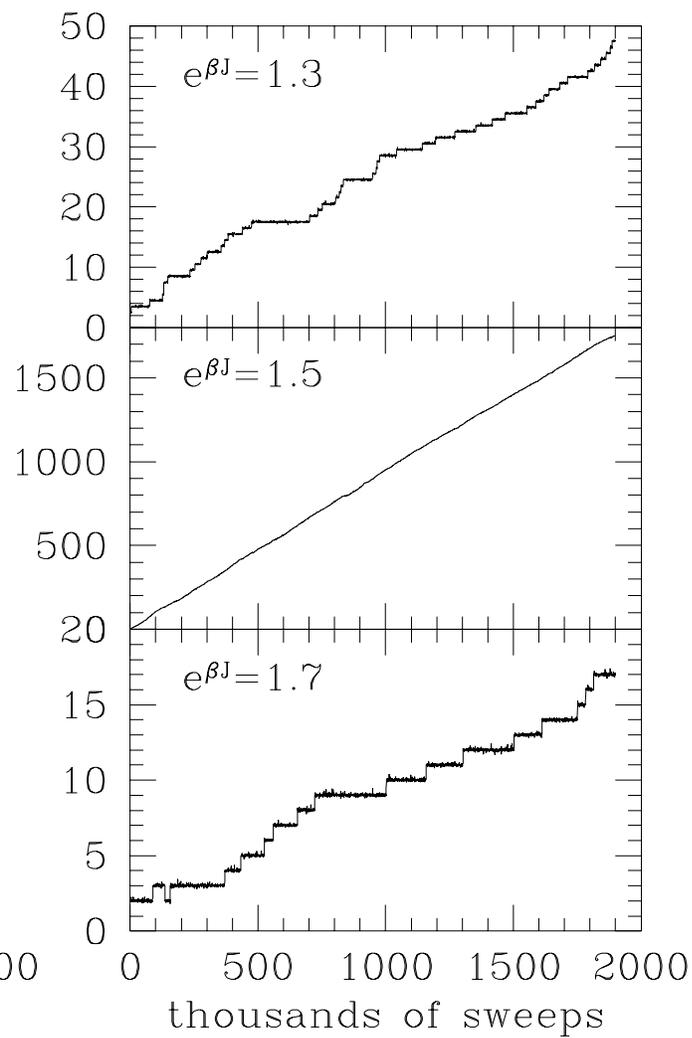
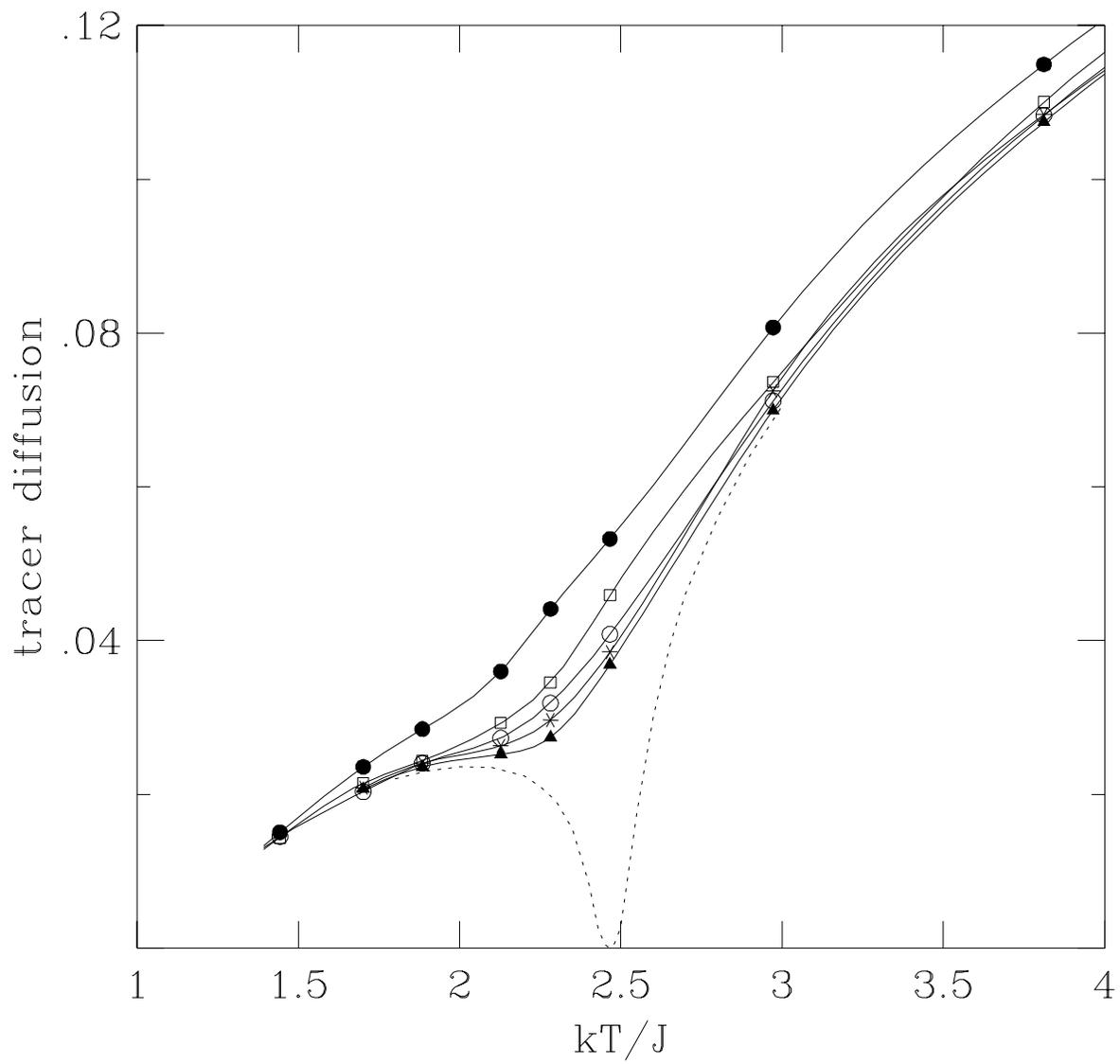
\end{document}